\documentclass[
 reprint,
 amsmath,amssymb,
 aps,superscriptaddress
]{revtex4-2}

\usepackage{graphicx}
\usepackage{dcolumn}
\usepackage{bm}
\usepackage{mathtools}
\usepackage{siunitx}
\usepackage{hyperref}

\DeclarePairedDelimiterXPP\BigOSI[2]
  {\mathcal{O}}{(}{)}{}
  {\SI{#1}{#2}}

\hypersetup{
  colorlinks=true,
  linkcolor=blue,
  filecolor=magenta,
  citecolor=blue,
  urlcolor=blue
}

\begin{document}

\preprint{APS/123-QED}

\title{Novel deep learning approach to detecting binary black hole mergers}

\author{Damon Beveridge}
 \email{damon.beveridge@research.uwa.edu.au}
\author{Alistair McLeod}
 \email{alistair.mcleod@research.uwa.edu.au}
\author{Linqing Wen}
 \email{linqing.wen@uwa.edu.au}
\affiliation{Department of Physics, The University of Western Australia, 35 Stirling Highway, Crawley, WA 6009, Australia
 }

\author{Andreas Wicenec}
\affiliation{
International Centre for Radio Astronomy Research, The University of Western Australia
M468, 35 Stirling Hwy,  Crawley, WA 6009, Australia
}

\date{\today}

\begin{abstract}
Gravitational wave detection has opened up new avenues for exploring and understanding some of the fundamental principles of the Universe. The optimal method for detecting modeled gravitational-wave events involves template-based matched filtering and performing a multidetector coincidence search in the resulting signal-to-noise ratio time series. In recent years, advancements in machine learning and deep learning have led to a flurry of research into using these techniques to replace matched filtering searches and for efficient and robust parameter estimation of the gravitational wave sources. This paper presents a feasibility study for a novel approach to detecting binary black hole gravitational wave signals, which utilizes deep learning techniques on the signal-to-noise ratio time series produced from matched filtering. We show that a deep-learning search can efficiently detect binary black hole gravitational waves from the signal-to-noise ratio time series in simulated Gaussian noise with simulated transient glitches. Furthermore, our search method can outperform a maximum SNR-based matched filtering search on simulated data of the Hanford and Livingston LIGO detectors in the presence of glitches. Lastly, since we are building upon the foundations of a matched filtering search pipeline, we can extract estimates for the signal-to-noise ratio and detector frame chirp mass of a gravitational wave event with similar accuracy as existing pipelines.
\end{abstract}

\maketitle

\section{\label{sec:intro}Introduction}

Gravitational waves (GWs) were first detected by the Laser Interferometer Gravitational-Wave Observatory (LIGO) \cite{LIGOScientific:2014pky} on September 14, 2015 \cite{Abbott:2016blz}. Since then, GW events from compact binary coalescences (CBCs) have been detected regularly across three separate observing runs \cite{gwtc1, gwtc2, gwtc21, gwtc3} by the LIGO and Virgo \cite{Acernese_2015} detectors, and the fourth observing run is now underway with the addition of the Kamioka Gravitational Wave Detector (KAGRA) \cite{KAGRA}. With upgrades to these existing detectors improving their sensitivity and additional detectors such as LIGO-India \cite{Saleem_2021} coming online, the opportunity to detect new GW events is increasing dramatically with each observing run. An increase in sensitivity brings an expected increase in the number of noise transients and complexity of gravitational wave data, making the development of sensitive search algorithms for the future of gravitational wave astronomy more important.

Currently, four CBC search pipelines operate in real-time to detect modeled gravitational-wave events. These pipelines are PyCBC Live \cite{Nitz2018, Dal_Canton_2021}, GstLAL \cite{Cannon2020, Ewing2024}, SPIIR \cite{QiChu2022} and MBTA \cite{Aubin_2021} and are all based on the technique of matched filtering, which employs a bank of modeled GW waveform templates to perform a cross-correlation with the detector strain data to identify matching signals. When the output from matched filtering achieves a predefined threshold, together with a signal consistency test \cite{Allen2005}, and no data quality issues arise \cite{Abbott_2018}, an event trigger can be produced. These tests, however, are not always robust in the presence of transient glitches in the detector noise and can still produce significant triggers without the presence of a gravitational wave \cite{Christensen_2004, gracedb}, or miss a detection due to the presence of a glitch \cite{gw170817}.

Deep learning \cite{deeplearning, Krizkhevsky2012, simonyan2015deep, Goodfellow-et-al-2016} is a subset of machine learning that uses artificial neural network models to make predictions by extracting features from input data for problems such as classification and regression. In recent years, machine learning and deep learning have been applied to gravitational-wave physics for problems such as binary black hole (BBH) merger detection \cite{George2018, George2018_Real, Huerta2021, Gabbard2018, Fan2019, Rebei2019, Gebhard2019, Santos2022, Corizzo2020, Lin2021, Deighan2021, WeiWei2021, Xia2021, Schafer2022, Schafer2022_2, Verma2022, Ma2022, Verma2022_2, Andrews2022, Yan2022, Barone2022, mlgwsc, Nousi2022, aframe, Trovato_2024, Zelenka2024, Yamamoto2023, Tian_2024, Chaturvedi2022, Sasaoka2024, Morales2021, Kim_2021, Alvares_2021, Menendez2021, Fan2021, Lopac2022, Ravichandran2023, Andres-Carcasona2023, Jadhav2021, Jadhav2023, Fernandes_2023, YXWang2023, Murali2023, Alhassan2022, Marianer2020, koloniari2024, Wang2020, Jiang2022, bresten2019detection, Choudhary2023}, glitch identification and classification \cite{Fernandes_2023, Powell2017, Powell2015, Mukund2017, Razzano2018, Cuoco2018, GravitySpy, Biswas2013, Colgan2020, iDQ, Cavaglia2018, Llorens-Monteagudo2018, Morawski2021, Bini_2023}, and parameter estimation \cite{Fan2019, Alvares_2021, Chatterjee2019, Mcleod2022, Chatterjee2022, Chatterjee2022_2, Dax2021, Yamamoto2023, Tang2024, Gramaxo_Freitas_2024, Langendorff2023}, with promising results.

Previous efforts on detecting gravitational waves from BBH mergers with deep learning has looked at making detections from the detector strain data \cite{George2018, George2018_Real, Huerta2021, Gabbard2018, Fan2019, Rebei2019, Gebhard2019, Santos2022, Corizzo2020, Lin2021, Deighan2021, WeiWei2021, Xia2021, Schafer2022, Schafer2022_2, Verma2022, Ma2022, Verma2022_2, Andrews2022, Yan2022, Barone2022, mlgwsc, Nousi2022, aframe, Trovato_2024, Zelenka2024, Yamamoto2023, Tian_2024, Chaturvedi2022, Sasaoka2024} or time-frequency spectrograms \cite{Sasaoka2024, Morales2021, Kim_2021, Alvares_2021, Menendez2021, Fan2021, Lopac2022, Ravichandran2023, Andres-Carcasona2023, Jadhav2021, Jadhav2023, Fernandes_2023, YXWang2023, Murali2023, Alhassan2022, Marianer2020, koloniari2024}. It has been demonstrated that deep learning models can reach a similar sensitivity or even outperform the current pipelines for BBH mergers with a source frame chirp mass (see Eq. \ref{eq:chirp_mass}) greater than $\sim$15$\,$M$_\odot$ \cite{mlgwsc, aframe}. However, detecting CBC mergers from these data sources using deep learning becomes difficult when looking at lower-mass events such as binary neutron star (BNS) and neutron star-black hole (NSBH) mergers, as well as low-mass BBH mergers \cite{mlgwsc, aframe}. This arises due to the GW signal duration increasing for lower-mass binaries, meaning the signal power is spread over a longer period, resulting in a lower overall amplitude in the detector strain for a given signal-to-noise ratio (SNR). There have also been publications highlighting efforts to detect BBH mergers that use the peak values of the SNR time series produced by matched filtering as input to a deep learning model, with small template banks consisting of 35 \cite{Wang2020} and 57 \cite{Jiang2022} template waveforms. An additional challenge for this area of research is developing a robust search method in the presence of nonstationary and non-Gaussian noise, especially those in the form of transient detector glitches \cite{mlgwsc, Choudhary2023, Jadhav2021, McIsaac_2022, Joshi_2021, Nitz_2018}.

This paper presents an alternative approach to detecting gravitational waves from binary black hole mergers using the SNR time series produced by matched filtering as the input to a deep learning model. This technique is advantageous in several aspects, including the fact that the SNR time series is readily available from current online and offline search pipelines, allowing an efficient implementation of this model to a current pipeline without additional data preprocessing and matched filtering computation. Another advantage is that the SNR time series projects the power of a gravitational wave signal into a small time window and is consistent for all CBC source types, allowing our deep learning model to be easily extended to detect lower mass mergers. Furthermore, matched filtering is the optimal search method for identifying modeled signals in stationary Gaussian noise \cite{wainstein1962}. However, data from the detectors can be non-stationary and contain transient glitches from environmental and instrumentation sources \cite{Abbott_2020, Davis_2021}. In this paper, we present a feasibility study using this approach to detect BBH signals. We use simulated Gaussian data from the Hanford (H1) and Livingston (L1) LIGO detectors with injected glitches to show that this deep learning approach can effectively detect BBH signals, especially in glitchy detector noise. Lastly, because we build our search method on top of matched filtering, we show how we can produce estimates for the SNR and detector frame chirp mass of a detected gravitational wave signal with similar accuracy to existing matched filtering search pipelines, despite selecting the trigger associated with these values based on the output of the model instead of the network SNR.

Section \ref{sec:method} discusses the methods we have used to generate data to train and test our deep learning model, the architecture of the deep learning model and the search pipeline setup used to make detections. Section \ref{sec:results} presents the results of our search method in terms of robust false alarm rate assignment, detection sensitivity, and parameter recovery. Lastly, Sec. \ref{sec:conc} summarizes our findings and presents further research that could be done in this area, as well as future applications of this work.

\section{\label{sec:method}Method}

\subsection{\label{sec:matched-filtering}Matched filtering}

Matched filtering is a fundamental technique in detecting gravitational waves and is crucial in identifying known signals buried in noisy detector data. It involves cross-correlating the detector strain, $s$, with a bank of template waveforms, $h$. The output of the matched filtering operation forms the signal-to-noise ratio (SNR) time series, $\rho(t)$, which can be used to estimate the likelihood of a GW signal being present in the data. A template waveform, $h(\theta)$, is defined by its intrinsic source parameters $\theta$, and the matched filtering operation is performed individually for each template waveform in the bank. The SNR time series is defined as \cite{shaunhooper, findchirp},

\begin{equation}\label{eq:1}
\rho(t) = \frac{z(t)}{\sigma},
\end{equation}
where $z(t)$ is the complex matched filter \cite{shaunhooper, findchirp}, defined as,

\begin{equation}\label{eq:2}
z(t)=2\int_{-\infty}^{\infty} \frac{\tilde{s}(f)\tilde{h}^*(f)}{S_n(|f|)}e^{2\pi ift}df.
\end{equation}
Here, $\tilde{s}(f)$ is the Fourier transform of the detector strain data, $\tilde{h}(f)$ is the Fourier transform of the template waveform, and $S_n(f)$ is the power spectral density (PSD) of the strain data. In Eq. \ref{eq:1}, $\sigma$ is the normalization constant computed by the square root of the inner product of the template with itself.

\begin{figure}[t]
\includegraphics[width=\linewidth]{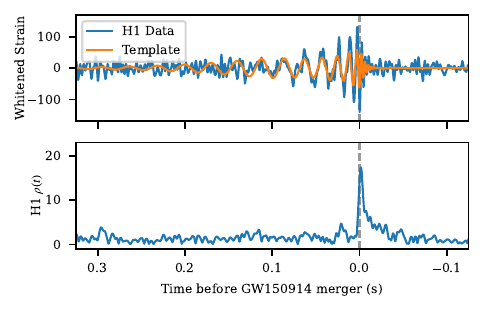}
\caption{\label{fig:strain-snr}Top panel: whitened LIGO Hanford (H1) detector strain data containing GW150914 (blue), and the optimal waveform template that is generated using the SEOBNRv4\_ROM approximant \cite{Bohe2017} and the median mass parameters from Ref. \cite{gwtc1} (orange). The waveform template is non-spinning and has been whitened and rescaled so that it visually overlaps with the signal in the strain data. Bottom panel: signal-to-noise ratio (SNR) time series for the event GW150914, from matched filtering using the strain data and waveform template shown in the top panel.}
\end{figure}

Figure \ref{fig:strain-snr} compares the detector strain and the SNR time series produced by matched filtering for the first real detected BBH GW event, GW150914. Here, the template waveform has the median component mass parameters estimated for the compact binary merger from Ref. \cite{gwtc1}. It can be seen that the signal power is much more concentrated in the SNR time series, making it a simpler task for a deep learning model. Note that lower binary masses can significantly increase the signal duration in the detector strain, while the power in the SNR time series would mostly remain within $0.1\,$s duration. The chirp mass is defined as

\begin{equation}\label{eq:chirp_mass}
\mathcal{M}_c = \frac{(m_1 m_2)^{3/5}}{(m_1 + m_2)^{1/5}}.
\end{equation}

In this work, we generate the templates for matched filtering using the SEOBNRv4\_ROM \cite{Bohe2017} waveform approximant with a low-frequency cutoff of $18\,$Hz. We used a subset of the template bank used by the online GstLAL pipeline in the O2 observation run \cite{gstlal_o2_bank}, which was subsequently used by the SPIIR pipeline in the O3 observation run. Our subset comprises all templates where the detector frame primary and secondary masses are within the bounds of $4-200\,$M$_\odot$, resulting in a bank of 19,365 templates. This space was chosen to target BBH signals with source frame masses of $5-100\,$M$_\odot$, to cover the edges of the parameter space, and to account for the cosmological redshift of higher mass sources.

\subsection{\label{sec:datasets}Training datasets}

In this work, we generate data for training and testing the deep learning model by generating Gaussian noise colored by the H1 and L1 LIGO detector PSDs. We also add simulated glitches that imitate real detector transient glitches to the noise background. For samples that contain a BBH signal injection, we project the signal onto the antenna pattern of each detector and inject them into the strain time series.

Our deep learning model makes predictions using the absolute value of the SNR time series from the H1 and L1 detectors as input. A PSD for each detector is calculated using the first seven days of strain data from the O3a observing run. The strain data used to generate these PSDs is sourced from the Gravitational Wave Open Science Center (GWOSC) \cite{O3aData}. Gaussian strain data for the training samples is generated from these PSDs using the PyCBC package \cite{pycbcpackage} at a sampling frequency of $2048\,$Hz. To avoid the filter wraparound issue of a discrete matched filter \cite{findchirp}, we generate noise samples with a length of 250 seconds, which is larger than the duration of the longest template waveform in our template bank, generated with the PyCBC package \cite{pycbcpackage}. We select a 1-second region in the latter half of the 250-second sample as the second of data that encompasses the training sample, and here we inject any gravitational wave signal or simulated glitch injections. This 1-second region is the second that gets saved as the SNR time series training sample after matched filtering.

We simulate and inject a random glitch to the simulated detector strain at random positions in the noise. This work uses a set of glitch injections modeled by Gaussian, sine-Gaussian, and ring-down signals, as used in Refs. \cite{Llorens-Monteagudo2018, Razzano2018, Mukund2017, Powell2015} to simulate glitches in detector strain data. Some works use simulated glitches with additional signal morphologies \cite{Razzano2018, Mukund2017, Skliris2024}; however, the signals used in this paper are selected based on their similar morphology to BBH GWs to enhance our deep learning model's detection ability. The Gaussian simulated glitches are defined by:
\begin{equation}\label{eq:gaussian}
h(t) = h_0 \times \text{exp}\left( -\frac{(t-t_0)^2}{2\tau^2} \right),
\end{equation}
which has an amplitude scaling factor $h_0$, a characteristic width $\tau$, and a peak time $t_0$. The sine-Gaussian glitches are represented as a sinusoidal signal with an amplitude that is modulated by a Gaussian,
\begin{equation}\label{eq:sinegaussian}
h(t) = h_0 \times \text{sin}(2\pi f_0 (t-t_0)) \times \text{exp}\left( -\frac{(t-t_0)^2}{2\tau^2} \right),
\end{equation}
where $f_0$ is the central frequency, and the characteristic width is derived from the central frequency and quality factor, $Q$:
\begin{equation}\label{eq:qualityfactor}
\tau = \frac{Q}{\sqrt{2}\pi f_0}.
\end{equation}
Lastly, the ring-down glitch type is modeled as a damped sinusoid, with input parameters matching the sine-Gaussian signal,
\begin{equation}\label{eq:ringdown}
h(t) = h_0 \times \text{sin}(2\pi f_0 (t-t_0)) \times \text{exp}\left( -\frac{(t-t_0)}{2\tau} \right).
\end{equation}

The bounds for each parameter used to generate these glitches can be found in Table \ref{tab:glitchparams}. All parameters are sampled uniformly within these bounds except for the amplitude scaling factor, which is sampled uniformly in base-10 log space. From large-scale tests, this parameter space results in a small set of glitches that can produce little to no effect on the SNR time series, up to peaks that reach an SNR of approximately 150.

{\renewcommand{\arraystretch}{1.25}%
\begin{table}[]
\begin{tabular}{cccc}
\hline \hline
Parameter   & Gaussian                      & Sine-Gaussian                 & Ring-down\\
\hline
$h_0$       & $[10^{-21}, 5\times10^{-21}]$ & $[10^{-21}, 5\times10^{-21}]$ & $[10^{-22}, 10^{-21}]$\\
$\tau$      & [0.0005, 0.005]               & -                             & -\\
$f_0$       & -                             & [40, 300]                     & [5, 20]\\
$Q$         & -                             & [40, 300]                     & [12, 20]\\
\hline \hline
\end{tabular}
\caption{\label{tab:glitchparams}Parameter ranges for injected glitches. The amplitude scaling factor, $h_0$, is sampled uniformly in base-10 log-space, and the other parameters are sampled uniformly in linear space.}
\end{table}}

The training dataset contains an equal number of samples containing a signal and samples that contain only noise. Additionally, in the training dataset, each detector has a 30\% chance of injecting a glitch into each strain sample, where the type of glitch is randomly selected and of equal probability. This chance is independent for each detector, meaning there is a 9\% chance of a random glitch occurring in both detectors for a given sample. The glitches are injected into the simulated noise before matched filtering. They are positioned with $t_0$ in the first 0.6 seconds of the 1-second time region that becomes the SNR time series input to the model. This accounts for the delay in the glitch response in the SNR time series due to the lengths of the templates.

We use an astrophysically distributed parameter space for the signal injections, similar to what was used in the offline analyses for GWTC-3 \cite{gwtc3}, and the parameter ranges and parameter priors for these injections can be seen in Table \ref{tab:params}. The component masses are sampled such that $m_1>m_2$, each sampled from a power-law distribution with powers of -2.35 and 1, respectively. In Table \ref{tab:params}, the mass ranges are presented in the source frame, and the component masses are projected into the detector frame based on the sampled extrinsic parameters. Aligned spins are used and sampled uniformly.

{\renewcommand{\arraystretch}{1.25}%
\begin{table}[]
\begin{tabular}{lccc}
\hline \hline
Parameter                       & Prior           & Minimum       & Maximum      \\
\hline
Primary Mass, $m_1$ (M$_\odot$)   & $m_1^{-2.35}$   & 5             & 100           \\
Secondary Mass, $m_2$ (M$_\odot$) & $m_2^{1}$       & 5             & $m_1$           \\
Component Spins, $S_z$          & Uniform           & -0.998        & 0.998        \\
Polarization Angle (rad)        & Uniform           & 0             & 2$\pi$       \\
Right Ascension (rad)           & Uniform           & 0             & 2$\pi$       \\
Declination (rad)               & Cosine            & -$\pi$/2      & $\pi$/2        \\
Inclination Angle (rad)         & Sine              & 0             & $\pi$        \\
Luminosity Distance (Mpc)       & Comoving          & 0             & 15000           \\
\hline \hline
\end{tabular}
\caption{\label{tab:params}Parameter ranges and sampling priors for simulated gravitational-wave injections. ``Comoving" refers to a uniform in comoving volume prior.}
\end{table}}

Using astrophysical distributions when sampling these injections results in a significant number of injections with an SNR so low that the signal cannot be recovered with the current detectors. To account for this, we discard any injections where the estimated network SNR is less than 6. To calculate this estimated SNR and to inject the signals into the simulated strain, we use the SEOBNRv4\_ROM waveform approximant \cite{Bohe2017} with a low-frequency cutoff of 18Hz. After generating the plus and cross polarizations of each injection, they are projected into the strain of the H1 and L1 detectors using the injection's sky position and orientation and the detector antenna patterns.

\begin{figure}[t]
\includegraphics[width=\linewidth]{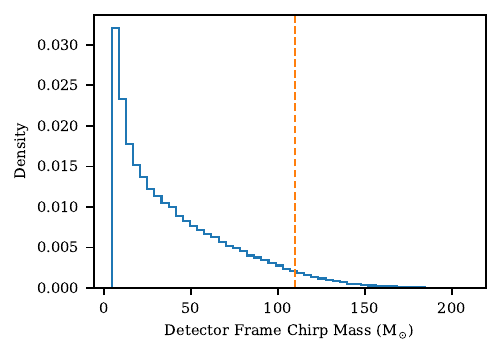}
\caption{\label{fig:mass-dist}Density histogram presenting the distribution of the detector frame chirp mass of signal injections in the training dataset after the minimum network SNR threshold is applied (blue). Only $5\%$ of injections in the dataset have a chirp mass higher than the orange dotted line, which sits at a chirp mass of $110\,$M$_\odot$. Our testing and analyses do not consider signals above this chirp mass, as our model has not been trained on enough samples in this region to expect good detection sensitivity.}
\end{figure}

Figure \ref{fig:mass-dist} shows the detector frame chirp mass distribution of this injection sampling for injections that pass the minimum network SNR threshold. There are noticeably fewer injections at high chirp masses, so we set a chirp mass limit for where we study our search method's sensitivity, as our training dataset will not have enough samples to train the deep learning model sufficiently. This limit is indicated in Fig. \ref{fig:mass-dist} as a chirp mass of $110\,$M$_\odot$, where only 5\% of injections exist with a higher chirp mass.

We sample 100,000 unique signal injections, each injected into unique strain samples. The dataset consists of 10 different SNR time series produced by different templates for each signal injection. To select the ten templates for each signal, we sort the templates in the template bank by chirp mass and extract the adjacent 250 templates on either side of the chirp mass of the injection. We compute the overlap of each template with the simulated injection waveform, weighted by the PSD of the H1 detector, as it is the least sensitive detector. A template is discarded if the resulting overlap is below 0.5. From the remaining set of templates, we select the template with the highest overlap to the simulated injection, and we randomly sample the nine remaining templates from those that passed the overlap cut-off. This results in a dataset of 1,000,000 SNR time series samples that contain signal injections. Additionally, the training dataset includes 1,000,000 SNR time series samples without any signal injections, which includes 100,000 unique strain samples, and we randomly sample 10 templates from the template bank for each strain sample. Of the 2,000,000 samples in the training dataset, 20\% are used for the validation dataset during training.

When generating each SNR time series of the samples containing an injection, the position of the merger time is randomly sampled to be in the central 0.9 seconds of the 1-second input window of the deep learning model. This is beneficial as it trains the deep learning model to search for patterns representing the existence of a GW signal across most of its input, which becomes crucial in constructing our ranking statistic in Sec. \ref{sec:pipeline}.

\subsection{\label{sec:ml}Deep learning model}

The architecture of our deep learning model is designed to identify the complex features present in the SNR time series in the presence of noise, signal injections and glitches. We treat this problem as a pattern recognition problem, where we attempt to detect the presence of a signal separately in each detector and then determine if the signal is coincident between the other detectors.

From our training dataset, the input to our model is two SNR time series samples of dimension $1\times2048$ each, one sample for each of the H1 and L1 detectors. The model output is a single prediction value indicating whether the input sample contains a gravitational wave signal. The model output uses a sigmoid activation function to constrain the output to between 0 and 1, where `1' represents a sample with a signal present.

\begin{figure}[h]
\includegraphics[width=\linewidth]{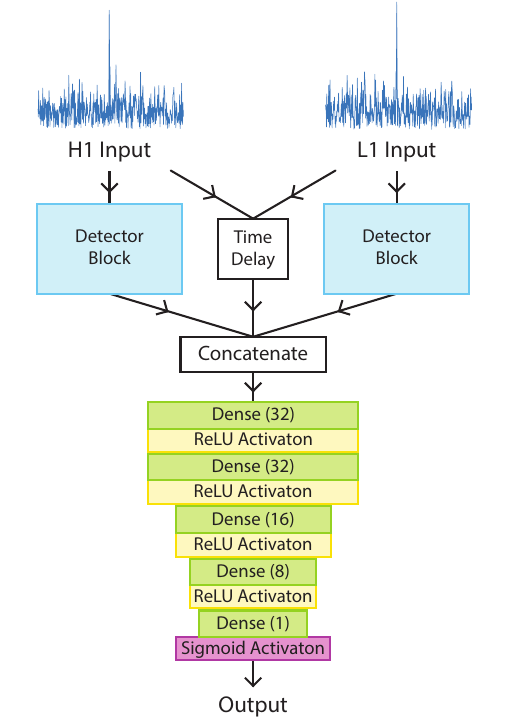}
\caption{\label{fig:model}Flow diagram of the overall model architecture. The model takes the SNR time series from the Hanford (H1) and Livingston (L1) detectors as inputs and feeds them into a detector block subarchitecture, which is depicted in Fig. \ref{fig:det_block}. The number in brackets for each dense layer is the number of units, or neurons, in that layer.}
\end{figure}

A broad overview of our deep learning model can be seen in Fig. \ref{fig:model}, which shows how we treat the detection problem in each detector separately using the detector blocks. The model concatenates the output of the two single-detector blocks and then uses a series of fully connected dense layers to assert if there is a coincident signal present. Additionally, we implement a `time-delay' layer which takes the two detector inputs and computes the time between the SNR peak of each sample in units of light travel time, where the light travel time is the maximum time separation of a coincident gravitational wave signal based on the distance between the H1 and L1 detectors. This layer is implemented to give the model additional confidence when detecting gravitational wave signals where the time-delay value is expected to be $\leq1$ unless a glitch is present or the signal is weaker than a noise peak.

\begin{figure}[h]
\includegraphics[width=\linewidth]{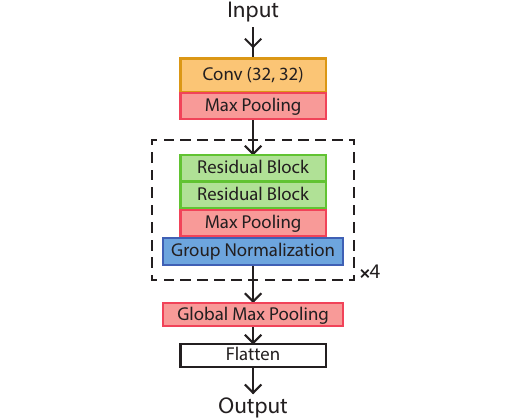}
\caption{\label{fig:det_block}Architecture of the detector blocks used in our deep learning model. The residual block subarchitecture can be found in Fig. \ref{fig:resnet}. The dotted box indicates that the block of layers is repeated four times. However, the final repetition does not include the max pooling layer due to the global max pooling layer used before the output. The orange block indicates a convolutional layer, with the values in brackets indicating the layer's number of filters and kernel size, respectively.}
\end{figure}

\begin{figure}[h]
\includegraphics[width=\linewidth]{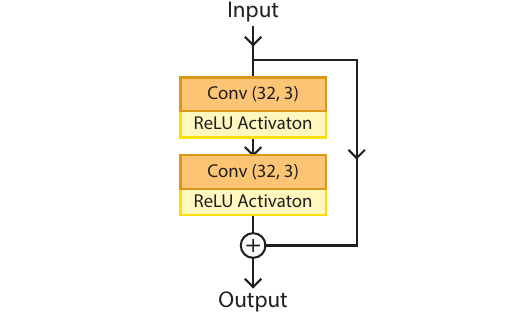}
\caption{\label{fig:resnet}Architecture of the residual neural network blocks used in the detector blocks of our deep learning model. The orange blocks indicate convolutional layers, with the values in brackets indicating the layer's number of filters and kernel size, respectively.}
\end{figure}

Each detector block has the same architecture, differing only by the weights and biases of each layer based on the data from each detector having separate PSDs, and this architecture can be seen in Fig. \ref{fig:det_block}. This block of layers implements convolutional neural network (CNN) layers \cite{cnn} to identify the patterns present in the data due to a gravitational wave signal. Combined with the CNN layers, we use a residual neural network (ResNet) structure \cite{he2015deep}, Fig. \ref{fig:resnet}, which implements a skip connection to prevent overfitting in a complex model. At the start of the detector blocks, the CNN layers have 32 filters and a kernel size of 32. The CNN layers in the ResNet layers also have 32 filters but have a kernel size of 3, use padding to make the output shape equal to the input shape, and include the rectified linear unit (ReLU) activation function \cite{relu}. The maximum pooling layers in the detector blocks have a pool size of 8 and a stride of 2. Motivated by \cite{aframe}, we utilize Group Normalization layers \cite{groupnorm} with a group size of 8 for normalizing across the channels, rather than the batch dimension, to aid in faster and more stable training. Lastly, we reduce the dimensionality of the output of the detector blocks using the flatten layer.

In the final stages of our model, it takes the outputs from the time-delay layer and the two detector blocks and concatenates them before feeding them through a series of dense layers. The purpose of these layers is to take the condensed representation from the two detector blocks, consider the time-delay information, and assign significance to the model output based on its inputs. We use a concatenation layer so that both detector block outputs have their complete encoded representation present and so that this stage of the model can learn the coincidence, or lack thereof, between the two detectors.

Our deep learning model was built, trained, and tested using the TensorFlow \cite{tensorflow2015} package. To train the model, we optimize by minimizing the binary cross-entropy loss function using the Adam optimizer \cite{kingma2017adam} with an initial learning rate of $10^{-3}$. The model training process reduces the learning rate by a factor of 10 if the loss does not decrease after 25 training epochs, down to a minimum learning rate of $10^{-6}$, and the model training is exited after 200 epochs. This training process takes 7 hours on a single NVIDIA A100 GPU.

Following training, we remove the sigmoid function from the model output and replace it with a linear function. This is motivated by some past work looking at removing bounds on the model's output for gravitational wave detection \cite{aframe, Schafer2022, Schafer2022_2, Andrews2022}, and because the model's float 32 precision results in predictions on many different inputs being rounded to 0 or 1 with the inclusion of the sigmoid function, having no significance between predictions of the same value. Replacing the sigmoid function with the linear function allows the model output to have an unbounded output to address this issue.

\subsection{\label{sec:pipeline}Search pipeline}

To go from our trained deep learning model to detecting gravitational wave events, we have set up a search method for producing triggers. Using this search method, we can define our ranking statistic and assign confidence to a trigger through a false alarm rate (FAR).

Our detection method uses a two-step process to produce a trigger on some data. The first step is to perform matched filtering on the strain time series from multiple detectors using every template in the template bank. This is followed by a coincident peak-finding search, which is aligned with how existing search pipelines produce triggers. The peak-finding search produces a network SNR, accounting for the light travel time between detectors for each template for every 1-second window in the data. At each second, we order the templates based on the network SNR. We also impose a minimum SNR of 4 in at least one detector so that we do not produce triggers when the SNR is too low for a detection to be made.

Since our model is the most accurate to templates with the highest signal response, we only make predictions with our model on the SNR time series from the ten highest network SNR templates for each 1-second window, based on the peak-finding outlined above. This also helps to manage the computational cost of this method, as the number of predictions will be orders of magnitude lower than if we made predictions over every template, which was found to be unnecessary. Since gravitational wave detector data can be considered as a continuous stream, we make a time series of 16 predictions per template instead of only one, as the peak identified can exist anywhere within the 1-second window, and our model will be less confident when peaks due to gravitational wave signals occur at the edges of its input window. These 16 predictions are performed at an inference rate of 16Hz, indicating they are overlapping predictions, and each one shifts across the SNR time series by $1/16$th of a second. Additionally, the 8th of the 16 predictions is centred at the peak SNR in the H1 detector for the produced trigger. For each of the ten templates involved in creating a trigger, we calculate the average of the 16 predictions. This allows us to reject short transient spikes in the model predictions and promote longer-duration high outputs from the model indicative of a signal existing in the input window.

The set of averaged predictions for each of the ten templates is then clustered down to a single ranking statistic value from a single template.  We select the template with the highest average prediction for our trigger, and this average prediction is used as our ranking statistic.

Using this method, we can perform analyses similar to current search pipelines. This includes background runs where we characterize our search method and deep learning model, allowing us to assign an FAR to a trigger, and injection runs where we can determine the model's sensitivity to detecting gravitational wave signals.

\section{\label{sec:results}Testing and Results}

In this section, we look at taking the search method that we have outlined in Sec. \ref{sec:method} and test its ability to have a stable ranking statistic, its sensitivity performance on a set of BBH injections, and how accurately it can produce SNR and chirp mass estimates of a BBH signal that it detects. To do this, we use a consistent simulated noise background and generate a set of astrophysical injections.

\subsection{\label{subsec:far_assign}False alarm rate assignment}

To properly analyze our search method's ability to detect BBH mergers, we first have to characterize our ranking statistic by collecting a background so that a false alarm rate can be assigned to triggers.

We prepare a dataset of $6\times10^6$ seconds, or $\sim$70 days, of simulated noise background with no injected gravitational wave signals to characterize our ranking statistic. Since our model is trained on simulated noise sampled from the PSDs of H1 and L1 from the first week of the O3a observing run, this new noise background uses the PSDs of each detector from the second week of O3a. The simulated glitch rate for this noise background is 0.8 per minute for H1 and 2.67 per minute for L1. This glitch rate is 2.67 times the median for the O3a observing run \cite{gwtc3}. This multiplier matches our glitch population with the glitch rate of a single-detector glitch with an SNR $\rho>6.5$ in the O3 observing run.

\begin{figure}[h]
\includegraphics[width=\linewidth]{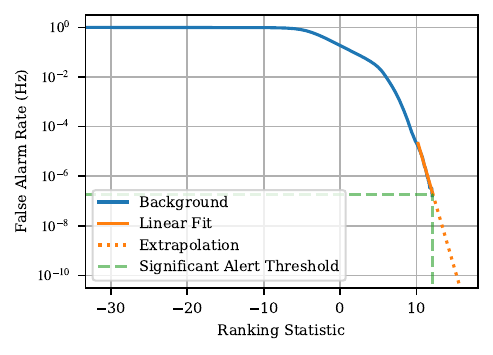}
\caption{\label{fig:far_extrap}Cumulative distribution of our ranking statistic (blue), plotted as a density to be able to assign false alarm rates. The solid orange line represents the linear fit on the tail of our background distribution, and the dotted orange line is the extrapolation of this fit used to assign lower false alarm rates. The green dotted line indicates the point at which the ranking statistic reaches a false alarm rate of 1 per 2 months based on the linear extrapolation.}
\end{figure}

We then use the search method outlined in \ref{sec:pipeline} to analyze this noise background and construct an estimated distribution of our ranking statistic on noise. Figure \ref{fig:far_extrap} presents a cumulative plot of our ranking statistic, which demonstrates how false alarm rates are assigned using the ranking statistic. Since our noise background consists of $6\times10^6$ seconds of data, we can only confidently assign false alarm rates to $1.67\times10^{-7}\,$Hz.

We implement a linear interpolation and extrapolation to our noise background to properly analyze our ability to detect gravitational wave signals, as shown in Fig. \ref{fig:far_extrap}. This conservative extrapolation is calculated by fitting a line to the points on the cumulative background below an FAR of $10^{-5}\,$Hz, and the fit has a coefficient of determination (R$^2$) of 0.982. We use the linear fit to assign an FAR when the ranking statistic is above ${\sim}11.45$, corresponding to an FAR of $10^{-6}\,$Hz. We interpolate FARs from this threshold as the number of background data points is small, resulting in significant step changes in FARs as the ranking statistic increases.

\begin{figure}[h]
\includegraphics[width=\linewidth]{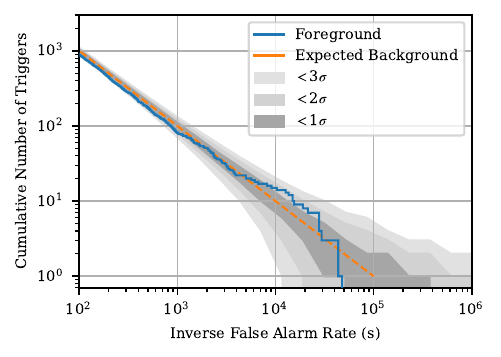}
\caption{\label{fig:ifar}Cumulative distribution of the inverse false alarm rate from triggers on Gaussian noise with simulated glitch injections. Our false alarm rate assignment defined on a collected background results in a distribution within 3$\sigma$ of the expected distribution.}
\end{figure}

Using the collected background for our ranking statistic and the method of assigning FARs to triggers, we can now test that it is robust to unseen noise. We do this by generating a new noise dataset using the PSDs from the third week of O3a to sample the detector strain and using the same glitch rate as the previous background. Figure \ref{fig:ifar} presents the cumulative counts of triggers for each inverse false alarm rate (IFAR). The triggers at each IFAR produced from this second analysis are within $3\sigma$ of the expected distribution.

\subsection{\label{subsec:inj_results}Injection results}

In this section, we present the detection performance of our search method on a set of injected gravitational wave signals. Sec. \ref{subsubsec:sens} indicates the general sensitivity of our search method over the parameter space we search for. In section \ref{subsubsec:missed-found}, we discuss the population of injections that we detect at an FAR of 1 per 2 months versus the population of injections that we fail to recover. Using these found injections, Sec. \ref{subsubsec:snr_recovery} presents how well our search method can recover the SNR of a population of signals. Lastly, Sec. \ref{subsubsec:cm_recovery} presents how our search method can make preliminary chirp mass estimates on each found injection based on the chirp mass of the template that was used to detect it.

\subsubsection{\label{subsubsec:sens}Sensitivity}

To analyze our search method's ability to detect simulated gravitational wave signals, we inject 10,000 BBH signals into a noise background. We then analyze the number of recovered signals at each FAR and the population of recovered signals at the significant alert threshold of 1 per 2 months. The injection set used in this testing matches the parameter distributions of the injection datasets used to analyze the sensitivity of the matched filtering search pipelines in Ref. \cite{gwtc3}. This differs from our training dataset by incorporating isotropic spins and reducing the component mass lower limit to $2\,$M$_\odot$. We use an upper spin magnitude limit of 0.998. When sampling injections, we incorporate a minimum estimated network SNR cutoff of 6, as in our training dataset, to remove any injections that are not expected to be recovered. We only perform tests on injections with both component masses $\geq 5\,$M$_\odot$ to match our training parameter space. As mentioned in Sec. \ref{sec:datasets}, only injections with a detector frame chirp mass $\leq110\,$M$_\odot$ are considered in our analysis due to the distribution of injections in our training dataset. Since these injections have isotropic spin components, we use the IMRPhenomPv2 \cite{imrphenomp} approximant for estimating the injection SNRs and SEOBNRv4PHMpseudoFourPN \cite{seobnrv4phm} for injecting the signals into the noise strain. The detector PSDs used for the noise background and for estimating each injection's SNR are the same week 3 PSDs used for the test that verified our FAR assignment, and we use the same glitch rate from this previous test.

For injection run tests, triggers are produced at one per second as outlined in Sec. \ref{sec:pipeline}. When injecting simulated gravitational wave signals, we randomly set the merger time so that it does not always occur at the same point of a 1-second window. Additionally, we generate new continuous noise strain samples for each injection so that signals do not overlap. When detecting an injected signal, we assign the trigger with the highest significance (lowest FAR) within $\pm1$ second of the injected merger time.

\begin{figure}[h]
\includegraphics[width=\linewidth]{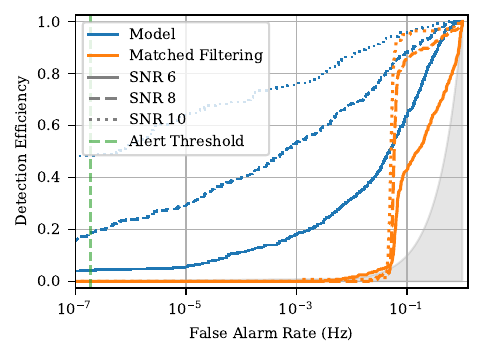}
\caption{\label{fig:roc_curve}Receiver operating characteristic curve indicating the relative sensitivity between our search method (blue) and one where the peak network SNR is used as a ranking statistic (orange). The injection samples used to calculate each curve are the injections from our test dataset within 0.1 of the estimated network SNRs in the legend. The shaded region indicates the performance of a random classifier on the test datasets, where the detection efficiency is equal to the false alarm rate.}
\end{figure}

The sensitivity of our search method can be seen in the receiver operating characteristic (ROC) curve of Fig. \ref{fig:roc_curve}. This plot focuses on our search method's ability to detect low network SNR signals. It compares this with a baseline search method that uses the maximum network SNR over the complete template bank as a ranking statistic. We plot the sensitivity of the two search methods over subsets of the total injection set, where the subset is all samples with an estimated network SNR in the range of $\pm0.1$ around a network SNR of 6, 8 and 10. We can see that due to the injection of simulated transient glitches, the sensitivity of the baseline matched filtering method drops off rapidly. In contrast, our search method demonstrates the deep learning model's ability to be robust in the presence of simulated transient glitches. It should be noted that when the baseline search curves drop rapidly, they follow the edge of the shaded region in Fig. \ref{fig:roc_curve}, which means it matches the sensitivity of a random classifier.

We see that our search recovers $34\%$ of all injections in a dataset that is heavily skewed toward injections with low network SNRs and low black hole masses. These initial sensitivity results are promising for applying this search method to detect BBH gravitational wave events, especially in the context of low network SNR signals. However, studies on real detector strain data and shared injection sets with existing search pipelines would be needed to confirm the feasibility of this search method.

Further optimizations could be explored for improving the search sensitivity in the future. These could include changing the training dataset parameter distributions, training loss and validation metrics, and new methods for the template selection of each sample in the training dataset.

\subsubsection{\label{subsubsec:missed-found}Missed and found injections}

\begin{figure}[h]
\includegraphics[width=\linewidth]{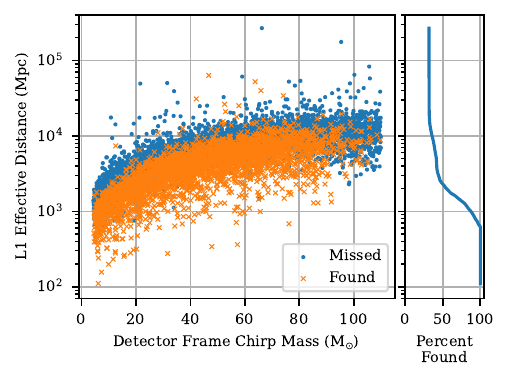}
\caption{\label{fig:missed_found}Missed and found injections from our test dataset. Injections are classified as found if the false alarm rate is less than 1 per 2 months.}
\end{figure}

The performance of our search method on the injection tests can be seen in Fig. \ref{fig:missed_found} in terms of missed and found injections against the Livingston detector's effective distance and chirp mass. We classify injections as found if the trigger associated with the injection has an FAR less than 1 per 2 months. Our search method can detect 100\% of BBH injections at an effective distance below $625\,$Mpc and more than 50\% of signals when the distance is below $2.2\,$Gpc. Our results are also consistent with the expectation that heavier BBH mergers can be detected further away.

Notable in Fig. \ref{fig:missed_found} is a small set of found injections at large effective distances in the Livingston detector relative to the general population of found injections. Intuitively, this would suggest that these injections have a low H1 effective distance and are being detected due to the higher SNR in the Hanford detector or because the injection occurs in a fortuitous noise realization where the Livingston SNR time series peaks above a low expected SNR value. Further analysis of these results indicates that some of these injections follow these expected explanations, however, there are still some outlier events in the relationship between effective distance and SNR. These remaining outliers are found to have high component spins, which results in spin-induced orbital precession. Because of this, the effective distance calculation that assumes a constant inclination is no longer valid, as the precession causes an inclination angle that varies with time. Further evidence of this can be seen in the GWTC-3 BBH injection set \cite{gwtc3_dataset}, which has the same priors as the injection dataset used in this work.

Figure \ref{fig:missed_found} also shows three missed injections at low effective distances where all other surrounding injections are classified as found. The missed injections at a chirp mass of $33\,$M$_\odot$ and $65\,$M$_\odot$, and effective distance of $1100\,$Mpc and $1500\,$Mpc, respectively, are missed due to the presence of loud glitches in one detector, and a low estimated SNR in the other detector, resulting in no opportunity for the model to make a confident detection. The other missed injection at a chirp mass of $20\,$M$_\odot$ and effective distance of $750\,$Mpc is missed by our search method as it is a single-detector event with an estimated SNR of 2 in H1 and 19.7 in L1, due to the source binary orientation and sky position. Not detecting this injection results from only considering a two-detector search method and not intentionally training our deep-learning model for this circumstance.

\subsubsection{\label{subsubsec:snr_recovery}SNR recovery}

\begin{figure}[h]
\includegraphics[width=\linewidth]{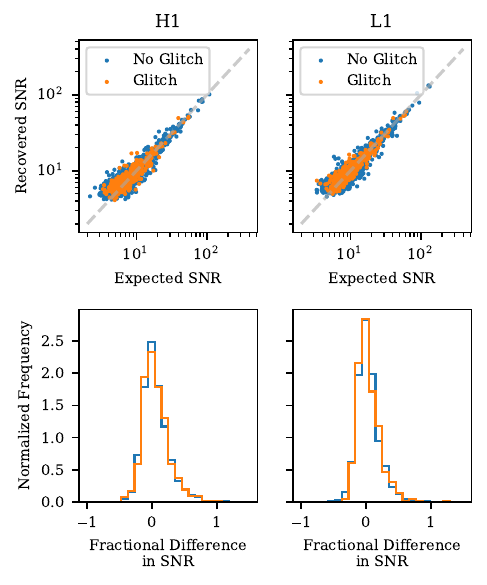}
\caption{\label{fig:rec_snr}Top panels: recovered and estimated SNRs for the found injections in our dataset in each interferometer: LIGO Hanford (H1) and LIGO Livingston (L1). The gray dashed lines indicate the diagonal where the recovered equals the estimated SNR. Bottom panels: fractional differences of the SNR in each detector, computed using Eq. \ref{eq:fractional}. The orange points are injections where one of the detectors contained a simulated glitch within $\pm$1 second of the merger.}
\end{figure}

Figure \ref{fig:rec_snr} demonstrates how accurately our search method recovers the SNR of a found injection in both the Hanford and Livingston LIGO detectors. The recovered SNR comes from the template's SNR time series, which produced the trigger. When sampling the signal injections, we compute the estimated SNRs in each detector by perfectly matching a projected injection waveform with itself using the same PSD as the noise background. Here, we have only included points where the H1 and L1 detectors are expected to recover an injection with a single-detector SNR greater than 4. The error, or fractional difference, of the recovered SNR is defined as

\begin{equation}\label{eq:fractional}
\text{error} = \frac{\text{recovered SNR} - \text{estimated SNR}}{\text{estimated SNR}}.
\end{equation}

We see in Fig. \ref{fig:rec_snr} that our search method generally recovers the estimated SNR in each detector, which reinforces the selection of our template bank despite it being designed initially for a different observing run PSD than what our noise background is sampled from. Additionally, there is more spread at low SNRs for the LIGO Hanford detector, which is likely caused by the fact that it is less sensitive than the LIGO Livingston detector. This results in injections having a typically lower SNR in H1 than L1, and results in low network SNR injections with a low estimated SNR in L1 having an even lower SNR in H1 and thus not being able to be found by our search method at the 1 per 2-month threshold.

\subsubsection{\label{subsubsec:cm_recovery}Chirp mass recovery}

\begin{figure}[h]
\includegraphics[width=\linewidth]{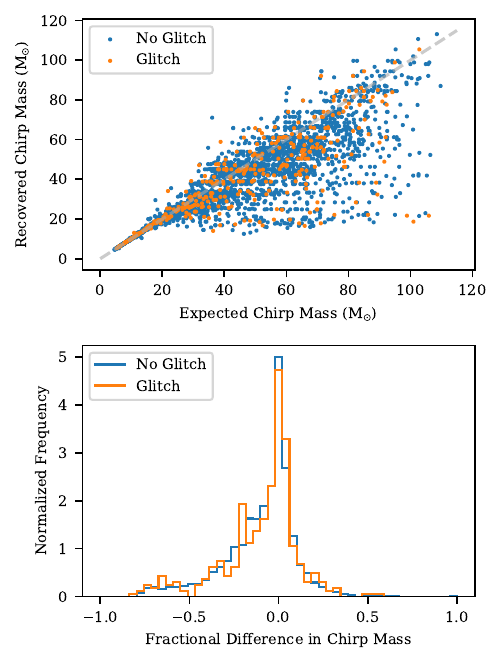}
\caption{\label{fig:rec_cm}Top panel: recovered vs expected detector frame chirp mass of the injections found by our search method. The gray dashed line indicates the diagonal where the recovered equals the expected chirp mass. Bottom panel: Fractional difference in the chirp mass computed in the same form as Eq. \ref{eq:fractional}. The recovered chirp mass is retrieved from the template that produces the trigger associated with an injection. The orange points are injections where one of the detectors contained a simulated glitch within $\pm$1 second of the merger.}
\end{figure}

Since our search method incorporates templates with matched filtering, we can make preliminary estimates for the detector frame chirp mass of a source based on the chirp mass of the template used to produce the trigger. We present the recovered chirp mass against the expected chirp mass for the found injections in Fig. \ref{fig:rec_cm}. The fractional difference in the recovered chirp mass is computed in the same form as Eq. \ref{eq:fractional}.

There is a notable spread in the recovered chirp mass for higher chirp mass injections, an expected feature of the BBH mass parameter space \cite{Ewing2024, QiChu2022}. This spread in recovered chirp mass for these higher-mass BBH signals is due to the low density of templates in this region of the template bank, as only a small number of templates are needed to search in this region sufficiently. This is due to these signals having a very short duration and a small frequency spectrum.

It is also clear from the fractional difference distributions in Fig. \ref{fig:rec_cm} that there is systematically no difference in the recovered chirp mass for injection times that include a glitch or not. This behavior can be due to several factors, including glitches existing in an adjacent trigger to the injection trigger, glitches having low strength, or glitches existing on the edge or outside the frequency spectrum of the template that produces the trigger.

\section{\label{sec:conc}Conclusion}

In this paper, we have presented the results of a two-detector deep learning search method that aims to detect a broad parameter space of binary black hole gravitational wave signals from the SNR time series produced in matched filtering. We have demonstrated that our search method is robust against three representative short-duration glitches in gravitational wave strain. Additionally, we have demonstrated a promising level of detection sensitivity, motivating future work. Lastly, we have shown that by incorporating matched filtering into our search method, we can produce SNR and chirp mass estimates for detected gravitational wave mergers with comparable accuracy to existing search pipelines, despite using the output of our model to select triggers.

Comparing our sensitivity to the existing matched filtering pipelines is limited due to the pipeline search results involving real detector noise. However, we have demonstrated the potential robustness of our search methods to non-stationary detector data by incorporating simulated transient glitches into our noise.

Following these results, the next logical step would be to apply this search method to real detector noise for a direct comparison with existing search pipelines and to have an opportunity to identify real gravitational wave events. Extensions of this work would include detecting more gravitational source types, like BNS and NSBH mergers, where current deep-learning approaches have limitations. Lastly, implementing this search method in an online, real-time detection scenario should be considered, as matched filtering search pipelines already exist, and we could efficiently access the SNR time series from them.

\begin{acknowledgments}

We wish to acknowledge the support of the Australian Research Council Centre of Excellence for Gravitational-Wave Discovery (OzGrav, Project No, CE170100004). This work was performed on the OzSTAR national facility at Swinburne University of Technology. The OzSTAR program receives funding in part from the Astronomy National Collaborative Research Infrastructure Strategy (NCRIS) allocation provided by the Australian Government, and from the Victorian Higher Education State Investment Fund (VHESIF) provided by the Victorian Government. This material is based upon work supported by NSF's LIGO Laboratory which is a major facility fully funded by the National Science Foundation.
\end{acknowledgments}

\bibliography{apssamp}

\end{document}